\documentclass[paper]{geophysics}
\pdfoutput=1
\usepackage{amsmath}
\usepackage{amssymb}
\usepackage{amsfonts}
\usepackage{color}
\usepackage{array}
\usepackage{seg}

\begin{document}

\renewcommand{\figdir}{.} 

\title{Multi-model full-waveform inversion}
\author{Musa Maharramov and Biondo Biondi}
\righthead{Multi-model FWI}
\lefthead{Maharramov and Biondi}
\footer{SEP--155}
\maketitle

\begin{abstract}
       We propose a multi-model formulation of full-waveform inversion that is similar to image decomposition into a ``cartoon'' and ``texture'' used in image processing. Inversion problem is formulated as unconstrained multi-norm optimization that can be solved using conventional iterative solvers. We demonstrate the proposed model decomposition approach by recovering a blocky subsurface seismic model from noisy data in time-lapse and single-model full-waveform inversion problems. 
\end{abstract}

\section{Introduction}
\cite{musasep1551} proposed a total-variation (TV) regularization technique for robust recovery of production-induced changes in the subsurface velocity model. \cite{musasep1552} proposed a TV-regularized constrained full-waveform inversion (CFWI) technique that achieves better constraining of the subsurface model in zones of poor illumination. In both cases the recovered model was assumed to be ``blocky''---i.e., contain areas of small (and predominantly monotone) velocity variation, as well as a few sharp contrasts along geologic interfaces. While sensible in many applications, ``blocky'' models represent an oversimplification of true physical properties. Even when the assumption of blocky behavior is justified, fitting noisy data may still result in an oscillatory model.

In this work we investigate splitting of a subsurface model into the sum of a ``blocky'' and ``wiggly'' components, and formulate a full-waveform inversion problem for recovering both components. Conceptually, our approach is similar to image decomposition into a ``cartoon'' and ``texture'' \cite[]{Meyer}, where ``texture'' is defined as a highly oscillatory pattern. We describe an implementation of multi-model full-waveform inversion of time-lapse and single-acquisition datasets, and demonstrate the method on two synthetic examples.

\section{Method}
A multi-model full-waveform inversion can be posed as an unconstrained regularized multiple-norm optimization problem:
                \begin{equation}
                        \begin{aligned}
                                \min\limits_{\mathbf{m},\mathbf{m}_b}\;  & \|\mathbf{F}(\mathbf{m})-\mathbf{d}\|^2_2\; +\; \alpha \| |\nabla \mathbf{m}_b |\|_1 + \beta \|\Delta \mathbf{m}_w\|^2_2,\\
 & \mathbf{m}\;=\;\mathbf{m}_b + \mathbf{m}_w,
                        \end{aligned}
\label{eq:mn}
                \end{equation}
where $\mathbf{m}_{b,w}$ are two \emph{uncorrelated} ``blocky'' and ``wiggly'' components of the model $\mathbf{m}$, with $\|\mathbf{m}_2\|_2 \ll \|\mathbf{m}_1\|_2$. The $\ell_1$ norm of the gradient, or the \emph{total variation} seminorm \cite[]{Triebel}, favors sharp contrasts over oscillations in $\mathbf{m}_b$, while the $\ell_2$ norm favors small oscillations over large contrasts. Solving (\ref{eq:mn}) produces a model split into two components, with one component exhibiting mostly blocky, and the other---oscillatory behavior. However, clean separation cannot be realistically achieved. Even in the simplest case of image TV denoising with the trivial modeling operator $\mathbf{F}(\mathbf{m})\equiv\mathbf{m}$, the two recovered components remain correlated \cite[]{Meyer,osher2005}. 

A problem analogous to (\ref{eq:mn}) can be formulated for time-lapse FWI,
\begin{equation} 
\begin{aligned}
 \min\limits_{\mathbf{m}_1,\mathbf{m}_2,\mathbf{m}_b}\;  & \|\mathbf{F}(\mathbf{m}_1)-\mathbf{d}_1\|^2_2\; + \|\mathbf{F}(\mathbf{m}_2)-\mathbf{d}_2\|^2_2\;+\\
 & \alpha \| |\nabla \mathbf{m}_b |\|_1 + \beta \|\Delta \mathbf{m}_w\|^2_2,\\
 & \mathbf{m}_2-\mathbf{m}_1\;=\;\mathbf{m}_b + \mathbf{m}_w,
\end{aligned}
\label{eq:4d}
\end{equation} 
where $\mathbf{m}_{1,2}$ and $\mathbf{d}_{1,2}$ are the baseline and monitor subsurface models and recorded data. 

We solve problems (\ref{eq:mn}) and (\ref{eq:4d}) by applying the nonlinear conjugate gradients algorithm \cite[]{Nocedal} after smoothing the TV  term,
\begin{equation}
\|| \nabla\mathbf{m} |\|_1 \;\approx\; \| \sqrt{|\nabla_{\mathbf{x}} m|^2+\epsilon} \|_1,
\label{eq:regtv}
\end{equation}
where $\epsilon\approx 10^{-5}$ is chosen as a threshold for realistic values of the slowness. Regularization parameters $\alpha$ and $\beta$ are chosen as follows. The value of $\alpha$ is chosen as in the standard TV-regularization problem with $\mathbf{m}=\mathbf{m}_b$ ($\mathbf{m}_2-\mathbf{m}_1=\mathbf{m}_b$ for time-lapse FWI) used in \cite[]{musasep1551}. Then $\beta$ is chosen sufficiently large to make oscillatory component close to zero, and gradually reduced until the models become correlated. 

Alternative constrained optimization problems can be formulated instead of (\ref{eq:mn}) and (\ref{eq:4d}), and solved using the approach described by \cite{musasep1552}. Alternatively, \emph{split-Bregman} of \cite{goldstein} can be applied directly to (\ref{eq:mn}) and (\ref{eq:4d}). 
\plot{truediff}{width=\columnwidth}
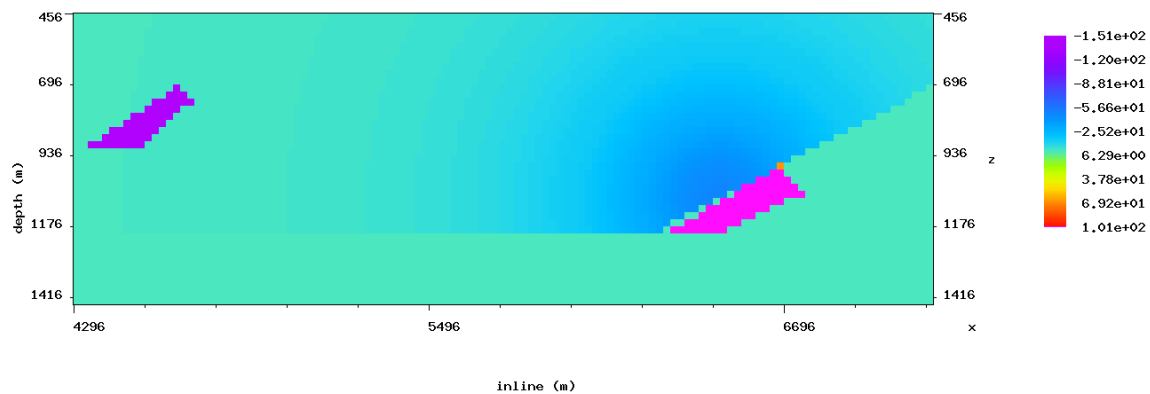
{True model difference showing two blocky anomalies (-150 m/s and 100 m/s) and a smooth velocity change peaking at -50 m/s in the overburden above the right anomaly.}

\section{Numerical examples}
We demonstrate multi-model inversion on the 7dB SNR Marmousi synthetic that we used in \cite[]{musasep1551,musasep1552}. See these papers for the details of numerical implementation.

\plot{pardiff}{width=\columnwidth}
{Model difference reconstructed using parallel difference algorithm \cite[]{Amir} for the 7dB SNR synthetic. Both amplitudes and locations of the anomalies are poorly resolved.}

For the time lapse example, a true model was generated consisting of ``blocky'' anomalies of -150 m/s and 100 m/s and a smooth velocity variation peaking at -50 m/s above the right anomaly (see Fig~\ref{fig:truediff}).

\plot{blocky}{width=\columnwidth}
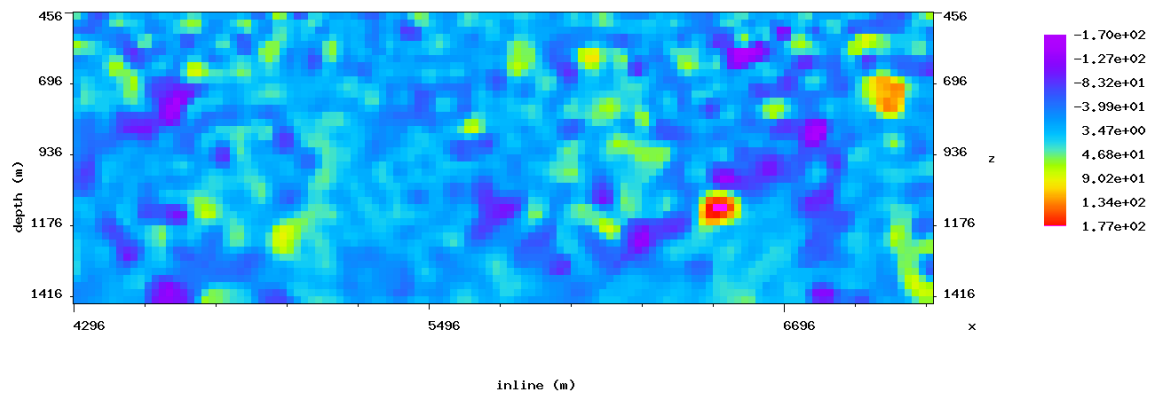
{Blocky component of the model difference recovered by solving (\ref{eq:4d}). The anomalies are resolved better, with amplitudes close to true values. The decomposition reveals partial recovery of the negative smooth velocity change over the right anomaly, however, most of the smooth velocity change ended up in the oscillatory component in Figure~\ref{fig:wiggly}. }

Note that the position of the smooth gradient and the small magnitude of velocity perturbations ($<5\%$ of baseline), in combination with noisy data, make resolution of the model difference very challenging in this case. 

\plot{wiggly}{width=\columnwidth}
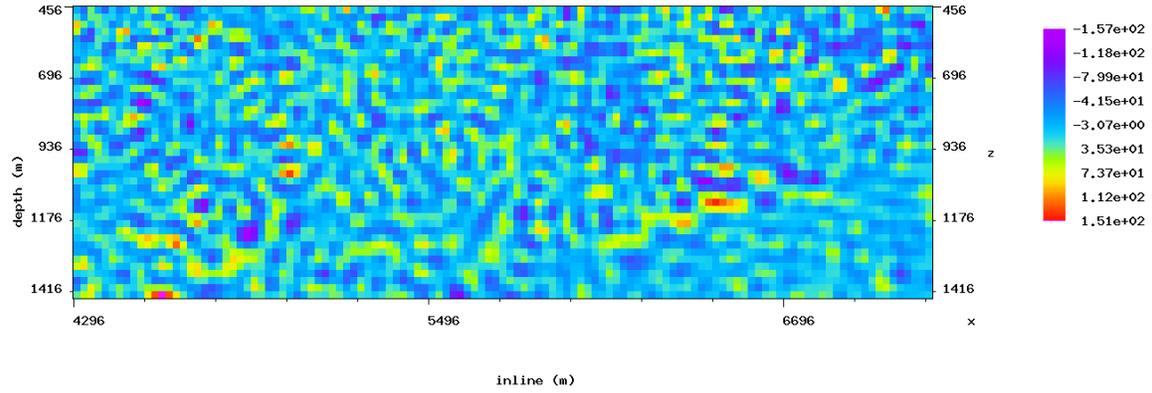
{Oscillatory component of the model difference recovered by solving (\ref{eq:4d}). Note that the two components appear to be mostly uncorrelated.}

This is confirmed by the result of parallel difference algorithm \cite[]{Amir} in Figure~\ref{fig:pardiff}. The anomalies are hard to identify, and their amplitudes are overestimated.

\plot{n4blocky}{width=\columnwidth}
{Blocky component of the baseline model recovered by solving (\ref{eq:4d}) for the 7dB SNR synthetic. }

The blocky component in Figure~\ref{fig:blocky} was recovered by solving (\ref{eq:4d}) with $\alpha=10^{-6}$ and $\beta=10^{-4}$. It appears to be a better approximation of the true model difference, both qualitatively and quantitatively. Note the partial recovery of the smooth velocity change in the overburden.

\plot{n4wiggly}{width=\columnwidth}
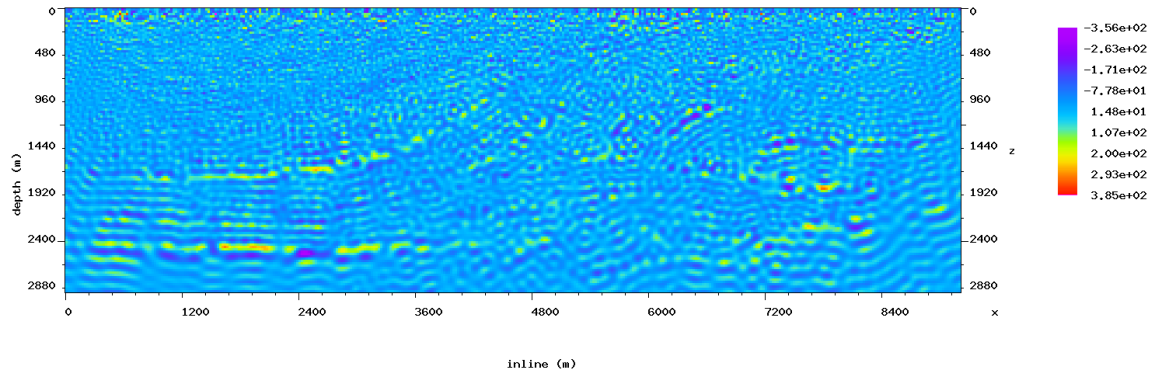
{Oscillatory component of the baseline model recovered by solving (\ref{eq:4d}) for the 7dB SNR synthetic. The model is largely uncorrelated with the blocky model of Figure~\ref{fig:blocky}, however, some of the sharp contrasts lave leaked into the wiggly model.}

The oscillatory component shown in Figure~\ref{fig:wiggly} is weakly correlated with the blocky model, and mostly represents noise removed from the model difference by he multiscale inversion (\ref{eq:4d}). Note, however, that the oscillatory component may contain some of the smooth overburden velocity change---see the negative velocity zone above the location of right anomaly in Figure~\ref{fig:wiggly}. A hierarchical multiscale decomposition approach similar to \cite[]{tadmor2004multiscale} may be applied to further decompose the model into blocky, \emph{smoothly varying} and oscillatory components.   

Figures~\ref{fig:n4blocky} and \ref{fig:n4wiggly} show the result of solving (\ref{eq:mn}) with the 7dB SNR synthetic. The two components demonstrate the typical pattern with multiscale multinorm decompositions: while the two blocky and oscillatory models are largely uncorrelated, some of the sharp contrasts have leaked into the wiggly model in Figure~\ref{fig:n4wiggly}.

\section{Conclusions}
Multi-model regularization of full-waveform inversion can be used for automated multiscale model decomposition. This can be exploited for isolating the effects of different physical processes acting on different scales, or separating useful information from the effects of fitting noisy data. The multi-model FWI can be implemented using the existing nonlinear unconstrained iterative solver frameworks with modest computational overhead, however, comparison with alternative methods \cite[]{goldstein,osher2005,cai,boyd} is necessary. Hierarchical model decompositions \cite[]{tadmor2004multiscale} and application to field data will be the subject of future work.

\section{Acknowledgments}
The authors would like to thank Jon Claerbout and Stewart Levin for a number of useful discussions, and the Stanford Center for Computational Earth and Environmental Sciences for providing computing resources.

\bibliographystyle{seg}  
\bibliography{mmbbmfwi}

\end{document}